\documentclass[aps,prl,10pt,reprint,superscriptaddress,showpacs]{revtex4-1} 

\usepackage{amsmath}     
\usepackage{amssymb}     
\usepackage{graphicx}    
\usepackage{hyperref}    
\usepackage{color}
\usepackage{subfigure}
\usepackage[utf8]{inputenc}

\newcommand{\ket}[1]{\ensuremath{| #1 \rangle}}
\newcommand{\half}{\ensuremath{\frac{1}{2}}}

\newcommand{\Cb}{\ensuremath{\bar{A}}}
\newcommand{\Db}{\ensuremath{\bar{B}}}
\newcommand{\cb}{\ensuremath{\bar{a}}}
\newcommand{\db}{\ensuremath{\bar{b}}}
\newcommand{\kvec}{\ensuremath{{\bf k}}}

\newcommand{\Th}{\ensuremath{\hat{T}}}
\newcommand{\dvec}{\ensuremath{{\bf d}_\kvec}}
\newcommand{\evec}{\ensuremath{{\bf e}}}

\newcommand{\sgn}{\ensuremath{\mbox{sgn}}}

\newcommand{\rvec}{\ensuremath{{\bf r}}}

\begin{document}

\title{Enhancing triplet superconductivity by the proximity 
to a singlet superconductor \\ in oxide heterostructures} 

\begin{abstract}
We show how in principle a coherent coupling between two 
superconductors of opposite parity can be realized in a three-layer 
oxide heterostructure. Due to strong intraionic spin-orbit coupling in the 
middle layer, singlet Cooper pairs are converted into triplet ones and vice 
versa. This results in a large enhancement of the triplet superconductivity, 
persisting well 
above the native triplet critical temperature. 
\end{abstract}

\author{Mats Horsdal}
\affiliation{Department of Physics, University of Oslo,
  P. O. Box 1048 Blindern, N-0316 Oslo, Norway} 
\affiliation{Institut f\"ur Theoretische Physik, Universit\"at Leipzig,
  D-04009 Leipzig, Germany} 
  
\author{Giniyat Khaliullin}
\affiliation{Max Planck Institute for Solid State Research, Heisenbergstrasse 1,
  D-70569 Stuttgart, Germany} 
  
\author{Timo Hyart}
\affiliation{Department of Physics and Nanoscience Center, University of Jyv{\"a}skyl{\"a}, P.O. Box 35 (YFL), FI-40014 University of Jyv{\"a}skyl{\"a}, Finland} 
\affiliation{Instituut-Lorentz, Universiteit Leiden, P.O. Box 9506, 2300 RA Leiden, The Netherlands}

\author{Bernd Rosenow}
\affiliation{Institut f\"ur Theoretische Physik, Universit\"at Leipzig,
  D-04009 Leipzig, Germany} 

\date{\today}


\maketitle

The prospect of realizing Majorana bound states that
can be used for quantum information processing has led to a large interest
in odd parity superconductivity. Native triplet superconductivity, 
believed to
be realized in, e.g., Sr$_2$RuO$_4$, is  
fragile and only present at very low
temperatures \cite{Mackenzie2003}. It is known that a singlet superconductor (SC) can induce triplet pairing
correlations in systems with Rashba spin-orbit coupling and/or ferromagnetism due to the proximity effect \cite{SigristUeda1991,BergeretEtal2001,Edelstein2003,EschrigLoefwander2008,TanakaSatoNagaosa2012,BergeretTokatly2013,ParhizgarBlack-Schaffer2014,VortexPhase,Lebed2006}, and it has been suggested to induce triplet superconductivity in hybrid structures 
with such properties~\cite{Sato09SO, Sau10,NakosaiTanakaNagaosa2012,Takei2013}. Rashba spin-orbit coupling and ferromagnetism can also give rise to a Josephson coupling between $s$ and $p$-wave SCs \cite{YangEtal2010,BrydonEtal2013}.

\begin{figure}
\centering
 \includegraphics[width=0.4\textwidth]{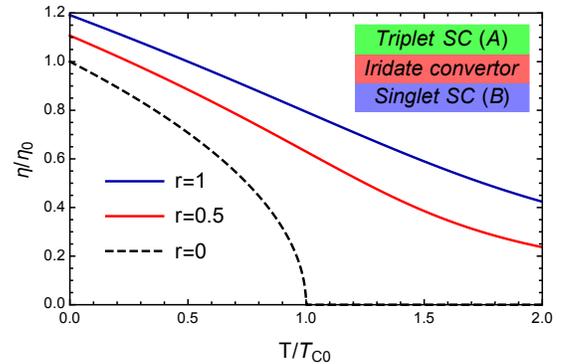}
\caption{(Color online) The $p$-wave order parameter $\eta$ as a function of temperature for different values of the enhancement parameter $r$ [defined below in Eq.~(\ref{eq:FreeEnergy})]. ($\eta_0$ and $T_{C0}$ are bare values of the order parameter and critical temperature). The inset shows the three-layer hybrid structure: a singlet SC (B), the "iridate convertor", and the triplet SC (A).} 
\label{fig:ProxEnhancement}
\end{figure}

Here we suggest an alternative way to improve the robustness of an odd 
parity SC by tunnel coupling it to an even parity singlet SC in all-oxide-based 
heterostructures. This mechanism is neither due to Rashba  
coupling nor ferromagnetism,  but by virtue of a strong {\it intraionic} spin-orbit 
coupling inherent to late transition metal compounds such as iridium oxide 
Sr$_2$IrO$_4$~\cite{BJKim}. To have a coherent coupling between two SCs of 
opposite parity, 
the tunneling has to "rotate" the Cooper pairs, since the wavefunctions 
of the odd and even parity superconducting condensates are orthogonal to 
each other. We consider a heterostructure consisting of three 
quasi two-dimensional layers: The even parity spin-singlet ''cuprate'' 
SC ($B$-layer) is separated from the odd parity spin-triplet
''ruthenate'' SC ($A$-layer) by an insulating layer, 
the ''iridate convertor'', see the inset of Fig.~\ref{fig:ProxEnhancement}. 
The superconductivity takes place in the $d_{x^2-y^2}$ band of the $B$-layer,
and in the Ru-orbitals of $t_{2g}$ symmetry in the $A$-layer. The tunnel 
coupling between the two SCs is provided via the spin-orbit entangled 
Ir $t_{2g}$-orbitals in the middle layer. We will show that a strong 
intraionic spin-orbit coupling in the middle layer gives rise to an
effective tunneling matrix between the two SCs, where the diagonal and
off-diagonal elements have opposite parity. The time reversal and
mirror symmetric tunneling matrix results in a coherent coupling between the
two SCs and leads to an enhancement of the odd parity order parameter, see Fig.~\ref{fig:ProxEnhancement}. 

This Rapid Communication is organized as follows: We start by describing the model
consisting of two SCs with opposite parity and show, on a
phenomenological level, that a general time-reversal and reflection-symmetric 
tunneling can give rise to coherent coupling of the order parameters of
different parity in the two layers. We then discuss experimental signatures 
of such a coupling, that is, a
dramatic enhancement of the triplet order parameter persisting 
well above the ''native'' critical temperature of the triplet SC. 
Finally, we show how the electron tunneling with the desired properties 
can be realized in oxide heterostructures, by considering a specific example 
of microscopic tunneling processes via the spin-orbit entangled Ir
$t_{2g}$-orbitals in the middle layer. 

\textit{Model}.---
The Hamiltonian of the system  
$H=H_A+H_B+H_{AB}$ 
comprises the Hamiltonians $H_{A/B}$ for the odd parity spin triplet $A$ layer 
and the even parity spin singlet $B$ layer, respectively, and the effective 
tunneling term $H_{AB}$ due to the "iridate convertor" between these layers. 
The Hamiltonian for the triplet-SC reads as $H_A=\half\sum_\kvec
\Cb_{\bf k}\bar{H}_A(\kvec)A_{\bf k}$, where the electron fields have been written in Nambu form, 
$\Cb_{\bf k}=(\cb_{{\bf k}\uparrow}, \cb_{{\bf k}\downarrow}, 
a_{-{\bf k}\uparrow}, a_{-{\bf k}\downarrow})$, and  
\begin{equation}
\bar{H}_A({\bf k}) = \left(
\begin{matrix}
\xi_A(\kvec) & \hat{\Delta}_A({\bf k}) \\
\hat{\Delta}_A^\dag({\bf k}) & -\xi_A({\kvec})
\end{matrix} \right).
\end{equation}
It is assumed that the dispersion $\xi_A({\bf k})$ is independent of spin. 
The order parameter matrix $\hat{\Delta}_A({\bf k}) =i \dvec\cdot
{\boldsymbol \sigma}\sigma_y$, where $\sigma_j$ ($j=x,y,z$) are the Pauli
matrices. 

The singlet-layer Hamiltonian $H_B$ takes a form similar to $H_A$, except for
the replacements $\xi_A({\bf k})\rightarrow\xi_B({\bf k})$, 
$\hat{\Delta}_A({\bf k})\rightarrow\hat{\Delta}_B({\bf k})$ with  
$\hat{\Delta}_B({\bf k}) =i \sigma_y \Delta_{B\kvec}$, and 
$\Cb_{\bf k}\rightarrow\Db_{\bf k}=(\db_{{\bf k}\uparrow}, \db_{{\bf k}\downarrow}, b_{-{\bf k}\uparrow}, b_{-{\bf k}\downarrow})$. 

A general tunneling term between the $A$ and $B$ layers can be written as
$H_{AB}=\half \sum_\kvec \Cb_{\bf k}T(\kvec)B_{\bf k} +H.c.$, where 
\begin{equation}
T(\kvec) = \left(
\begin{matrix}
\hat{T}(\kvec) & 0 \\
0 & -\hat{T}^*(-\kvec)
\end{matrix} \right), \quad
\hat{T}(\kvec) = \left(
\begin{matrix}
P_\kvec & R_\kvec \\
S_\kvec & Q_\kvec
\end{matrix} \right).
\label{eq:Tmatrix}
\end{equation}
Time reversal invariance of the Hamiltonian gives the following restriction on
the elements of the tunneling matrix: $P_\kvec=Q^*_{-\kvec}$ and
$R_\kvec=-S^*_{-\kvec}$. 
The system under consideration is invariant under reflections about the
$xz$-plane, $\mathcal{M}_x$, where the position and spin transform as
$(x,y)\rightarrow(x,-y)$ and $(S_x,S_y,S_z)\rightarrow(-S_x,S_y,-S_z)$. There is a similar symmetry under reflection about the
$yz$-plane, $\mathcal{M}_y$, so that in the spin sector $\mathcal{M}_x$ and $\mathcal{M}_y$ corresponds to $i\sigma_y$ and $i\sigma_x$, respectively. Since a spin-orbit coupling  
$\mathbf{L}\cdot\mathbf{S}$ is invariant under these symmetries, the tunnel Hamiltonian $\Th$ is invariant under the combined operation $\mathcal{M}_x\mathcal{M}_y$ and obeys $\sigma_z\Th(-\kvec)\sigma_z=\Th(\kvec)$. 
For this reason $P$ and $Q$ are even, $P_{-\kvec}=P_\kvec$ and $Q_{-\kvec}=Q_\kvec$, and $R$ and $S$ are odd, $R_{-\kvec}=-R_\kvec$ and $S_{-\kvec}=-S_\kvec$ \cite{RashbaSOC}.

The free energy for the system can be calculated by
integrating out the fermionic degrees of freedom~\cite{AltlandSimons}. It
takes the form $F=F_A+F_B+F_{AB}$, where $F_{A(B)}$ 
is the free energy for the $A(B)$ layer  
and $F_{AB}$ is the coupling between the two
layers. Our focus is on the coupling between the two SCs. Assuming time
reversal invariance of the system and unitary $p$-wave superconductivity in the $A$-layer, the coupling to second order in $H_{AB}$ is~\cite{SigristUeda1991}: 
\begin{eqnarray}
F_{AB}&\simeq& \half \sum_\kvec W_\kvec  \Delta^*_{B\kvec} 
\big\{d_{z\kvec}(|P_{-\kvec}|^2+|R_{-\kvec}|^2) \\
&+&(d_{x\kvec} \!+\! id_{y\kvec}) P_\kvec R_{-\kvec}
  +(d_{x\kvec} \!-\! id_{y\kvec}) P^*_\kvec R^*_{-\kvec}\big\} + H.c. , \nonumber
\label{eq:FPD}
\end{eqnarray}
where only the terms sensitive to the phase difference between the two layers
have been kept \cite{FNHigherOrder}. 
Here $W_\kvec =(\Phi^A_\kvec -\Phi^B_\kvec)/(E^2_{B\kvec}-E^2_{A\kvec})$, a function 
$\Phi^{A/B}_\kvec=\frac{1}{E_{A/B}}\tanh\frac{\beta E_{A/B}}{2}$, 
and $\beta$ is the inverse temperature. The quasiparticle energies are given by
$E_{A/B}(\kvec)=\sqrt{\xi^2_{A/B}(\kvec)+|\Delta_{A/B\,\kvec}|^2}$.
The symmetry $\mathcal{M}_x\mathcal{M}_y$ corresponds to $\kvec\rightarrow
-\kvec$ and a $\pi$ rotation of spins around the $z$-axis, such that the
invariance of the $\dvec$-vector implies $d_{x/y\,-\kvec}=-d_{x/y\,\kvec}$ and
$d_{z\,-\kvec}=d_{z\kvec}$, and as a consequence, $d_{z\kvec}\equiv0$. 
In Eq.~(\ref{eq:FPD}), this is reflected by the fact that 
$d_{x/y\,\kvec}$ ($d_{z\kvec}$) is
multiplied by an odd (even) functions of $\kvec$, 
so only $d_{x/y\,\kvec}$ terms may couple to a $\kvec$-even 
$\Delta_{B\kvec}$. 

To illustrate
the effect with a simple toy model, we assume that the
elements of the tunneling matrix take the simple 
form $P_\kvec=iP$ and $R_\kvec=R(\sin k_x -i\sin k_y)$, 
with $P$ and $R$ real. 
To get an idea which combinations of order parameters in the singlet and
triplet layers give a non-vanishing coupling, we consider the
following  
order parameters for the triplet-layer~\cite{SigristUeda1991} 
\begin{align}
\begin{split}
\Gamma^-_{1/3}: \quad &\dvec= \eta e^{i\theta}(\evec_x\sin k_x \pm \evec_y\sin k_y) \\
\Gamma^-_{2/4}: \quad &\dvec= \eta e^{i\theta}(\evec_x\sin k_y \mp \evec_y\sin k_x), 
\end{split}
\label{eq:pwavelist}
\end{align}
and that the singlet layer pairing has either $s$ or $d$-wave symmetry, 
$\Delta^s_{B\kvec}=\Delta_0(\cos k_x+\cos k_y)$ and 
$\Delta^d_{B\kvec}=\Delta_0(\cos k_x-\cos k_y)$, respectively. 
Here $\eta, \Delta_0>0$, and $\theta$ is the phase difference between the $p$- and $s/d$-wave order 
parameters. For this model, except
for the cases $\Gamma^-_2$ and $s$-wave or $\Gamma^-_4$ and $d$-wave, the integrand in Eq.~(\ref{eq:FPD}) is either exactly vanishing, antisymmetric under mirroring about the $y$-axis, or
antisymmetric under a $90^\circ$ rotation about the origin. Therefore only the combinations
$(\Delta^s_{B\kvec},\Gamma^-_2)$ and $(\Delta^d_{B\kvec},\Gamma^-_4)$ 
give a nonvanishing $F_{AB}$.
The first combination constitutes a fully gapped helical topologically non-trivial SC \cite{Sato2009, QiEtAl2010, FuBerg2010}. 

\textit{Proximity enhancement}.---The
coupling between the two SCs leads to a dramatic enhancement of the
triplet order parameter as we will see now. Close to the native critical
temperature $T_{C0}$ of the $A$-layer, the triplet order parameter is small 
and the free energy can be expanded in $\eta$. Assuming that the singlet 
pairing in the $B$-layer is robust, $\Delta_0 \gg T_{C0}$, and its variation 
near $T_{C0}$ is negligible, we can ignore the $F_B$ term since it only contributes a constant to the energy. To fourth order in $\eta$, the free energy can then be written as  
\begin{equation}
F=(t-1)a\eta^2 + \half b\eta^4 -ra\eta_0\eta\cos\theta,
\label{eq:FreeEnergy}
\end{equation}
where $a,b>0$ are constants describing the native $A$-layer \cite{Leggett1975, XuEtal1998}, and $t=T/T_{C0}$. The enhancement factor $r$ is defined by $\frac{F_{AB}}{a\eta_0}=-r\eta\cos\theta$. Here, $\eta_0$ is the zero temperature gap at vanishing $F_{AB}$. 
For fixed $\eta$,
it is clear that $F$ is minimized when $\cos\theta=\sgn(r)$. The
value attained by the $p$-wave order parameter is found by minimizing $F$ with
respect to $\eta$. Figure~\ref{fig:ProxEnhancement} shows $\eta$ as a function
of temperature for representative 
values of $r$ (see the discussion below). We see that the coupling to
the $B$-layer can give a large enhancement of the triplet superconductivity 
persisting well above $T_{C0}$. 
\begin{figure}
\centering
 \includegraphics[width=0.4\textwidth]{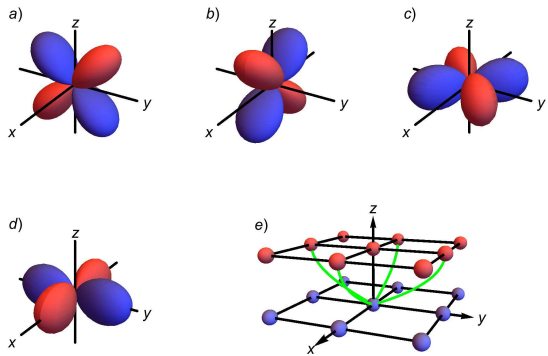}
\caption{(Color online) The shape of the atomic orbitals involved: 
a), b) $d_{yz}$ and $d_{xz}$ orbitals for the iridate convertor, c) $d_{xy}$-orbital for the iridate convertor and $A$-layer, and d) $d_{x^2-y^2}$-orbital for the $B$-layer.
e) The hopping paths (green lines) between the $B$-layer (bottom) and the iridate convertor layer (top). Similar paths apply to the hopping between the iridate convertor layer and the $A$-layer.
} 
\label{fig:OrbsLattice}
\end{figure}

Due to the amplification of the triplet order parameter, the
anomalous pair-tunneling current will persist also above the ''native''
critical temperature $T_{C0}$ in the form of a zero-bias peak in the
differential conductance \cite{KadinGoldman1982}. 
More directly, the proximity enhanced triplet gap (proportional to $\eta$) and its
temperature dependence (see Fig.\ref{fig:ProxEnhancement}) can be probed by scanning tunneling microscopy (STM) or angle-resolved photoemission spectroscopy (ARPES)
experiments.

\textit{Iridate convertor}.---We now discuss a possible realization of
the layered structure that gives rise to a tunneling matrix of the form
(\ref{eq:Tmatrix}), where the diagonal elements have even parity and the
off-diagonal ones have odd parity, providing a coherent coupling between
the $A$ and $B$ SCs.

We assume that all three layers have a square lattice geometry with similar 
lattice constants. A possible candidate, which can be designed by a modern
layer-by-layer growth technique~\cite{Hwang}, could be the oxide 
heterostructure Sr$_2$RuO$_4$/Sr$_2$IrO$_4$/La$_2$CuO$_4$. The pairing
in the $B$ (cuprate) layer takes place in the Cu $d_{x^2-y^2}$ orbitals 
designated by the annihilation operator $b_{\rvec\sigma}$, where $\rvec$ is 
the (two-dimensional) lattice position and $\sigma$ the spin, while the 
pairing in the $A$ (ruthenate) layer is assumed to take place in the 
Ru $d_{xy}$ orbitals~\cite{footnote1}, labeled by $a_{\rvec\sigma}$. 
The relevant orbitals in the middle layer are the 
Ir $t_{2g}$ orbitals $d_{yz}$, $d_{xz}$, $d_{xy}$   
denoted below by $\alpha_{\rvec\sigma}$, $\beta_{\rvec\sigma}$,
$\gamma_{\rvec\sigma}$, 
respectively. Figure~\ref{fig:OrbsLattice} shows the orientation of the above orbitals. 

We consider first the tunneling between the $B$-layer and the
middle layer. Figure~\ref{fig:OrbsLattice} shows the possible hopping paths between the two layers.
There can be no hopping between a $d_{x^2-y^2}$ orbital
located at $\rvec$ and a $d_{xz}$ orbital located at $\rvec\pm\evec_y+\evec_z$
for symmetry reasons~\cite{SlaterKoster1954}. 
On the other hand, hopping from a $d_{x^2-y^2}$ orbital located at $\rvec$ to a $d_{xz}$ orbital
located at $\rvec\pm\evec_x+\evec_z$ is symmetry allowed. However there will be a relative sign difference between hopping
in the positive and negative $x$-directions. 
A similar argument applies to
hopping between a $d_{x^2-y^2}$ orbital in the $B$-layer and a $d_{yz}$ orbital
in the middle layer. 
The tunneling between the $B$-layer $d_{x^2-y^2}$ orbitals and the $d_{xz}$ and $d_{yz}$ orbitals
in the middle layer is then
\begin{equation}
tb^\dag_{\rvec\sigma}\left[
(\alpha_{\rvec-\evec_y}-\alpha_{\rvec+\evec_y})-
(\beta_{\rvec-\evec_x}-\beta_{\rvec+\evec_x})
\right]_\sigma+H.c.,
\end{equation}
where $t$ is the hopping strength. The
relative sign difference for hopping in the opposite $x$($y$)-direction will give
rise to the odd parity elements in the tunneling matrix. Due to the relative  $45^\circ$ rotation of the Cu $d_{x^2-y^2}$ and Ir $d_{xy}$ orbitals,  
there will 
always be equal contributions of opposite sign in an overlap
integral~\cite{SlaterKoster1954}. The same argument gives a vanishing element
for hopping in the straight $\evec_z$-direction~\cite{footnote2}. 

We recall now that the spin and orbital states of the Ir-ion are strongly 
entangled via intraionic spin-orbit coupling, 
$H_{SO}=\lambda\mathbf{L}\cdot\mathbf{S}$, which results in a completely 
filled $J_{eff}=3/2$ quartet well below the Fermi level, and a half-filled 
$J_{eff}=1/2$ doublet~\cite{BJKim}. This implies that the tunneling 
process is mostly contributed by half-filled $J_{eff}=1/2$ states, 
with the following wave-functions~\cite{BJKim,Khaliullin}:          
\begin{equation}
\ket{f_{\sigma}}=\frac{1}{\sqrt{3}}\left\{ \sigma\ket{yz,-\sigma}
+i\ket{xz,-\sigma}+\ket{xy,\sigma}\right\}. 
\end{equation}
Projection of the Ir $t_{2g}$ states onto the $f_{\sigma}$ band gives the
following correspondence: 
\begin{equation}
(\alpha_{\rvec,\sigma};\; \beta_{\rvec,\sigma};\; \gamma_{\rvec,\sigma})\rightarrow
\frac{1}{\sqrt{3}} (-\sigma f_{\rvec,-\sigma};\; i f_{\rvec,-\sigma};\; f_{\rvec,\sigma}). 
\end{equation}  
With this substitution in Eq.~(7) and after a Fourier transformation, we 
arrive at the tunneling between the $B$-layer and the middle (M) layer:  
\begin{equation}
H_{BM}=\frac{2}{\sqrt{3}}\sum_{\kvec\sigma} t(\sin k_x-i\sigma \sin k_y)
b^\dag_{\kvec\sigma} f_{\kvec-\sigma} +H.c. 
\end{equation}

We now consider the tunneling between $t_{2g}$  
orbitals in the middle layer and the $d_{xy}$-orbital in the $A$-layer. By
arguing as above, we find that there will only be hopping between
$d_{xz}$ orbitals at $\rvec$ and $d_{xy}$ orbitals at
$\rvec\pm\evec_y+\evec_z$; we denote this hopping by $t'$. Similar arguments 
apply to the hopping $t'$ between $d_{yz}$ orbitals and $d_{xy}$ orbitals.  
There is also a hopping between the $d_{xy}$ orbitals of the middle and
$A$ layer. In this case there is no relative minus sign for hopping in
the opposite $x(y)$-direction and hopping to all next-nearest neighbors have the
same magnitude, which gives a tunneling of the form
$t''\gamma_{\rvec\sigma}^\dag(a_{\rvec\pm\evec_x}+a_{\rvec\pm\evec_y})_\sigma +H.c.$. 
Projecting the above $t'$ and $t''$ hopping processes onto the $J_{eff}=1/2$ band, we find 
the tunneling Hamiltonian between the middle layer and the $A$ layer:  
\begin{align}
H_{MA}=&\frac{2}{\sqrt{3}}\sum_{\kvec\sigma}f^\dag_{\kvec\sigma}
\big[- it'\sigma(\sin k_x-i\sigma \sin k_y)a_{\kvec-\sigma} \nonumber \\
&+t'' (\cos k_x+\cos k_y)a_{\kvec\sigma}\big]+H.c.
\end{align}

Introducing an effective charge transfer energy $\Delta E$ required to move  
an electron into the middle layer, we can calculate the effective tunneling 
Hamiltonian between the Cu $d_{x^2-y^2}$ orbitals and the Ru $d_{xy}$ orbitals 
to second order in $H_{BM}$ and $H_{MA}$. We then find that the elements of 
the tunneling matrix (\ref{eq:Tmatrix}) are given by 
\begin{equation}
\begin{split}
P_\kvec& = ig_e (\sin^2 k_x+\sin^2 k_y), \\
R_\kvec& = -g_o (\cos k_x+\cos k_y)(\sin k_x -i\sin k_y),
\end{split}
\label{eq:PQRS}
\end{equation}
with amplitudes $g_e=\frac{4}{3\Delta E}tt'$ and $g_o=\frac{4}{3\Delta E}tt''$
in even- and odd-parity channels, correspondingly. While it is difficult to quantify these constants, the above orbital-symmetry considerations confirm that the desired topology of the tunneling matrix, with an opposite parity of the diagonal and off-diagonal elements, is indeed realistic in perovskite-type oxide heterostructures.    

\begin{figure}
\includegraphics[width=0.4\textwidth]{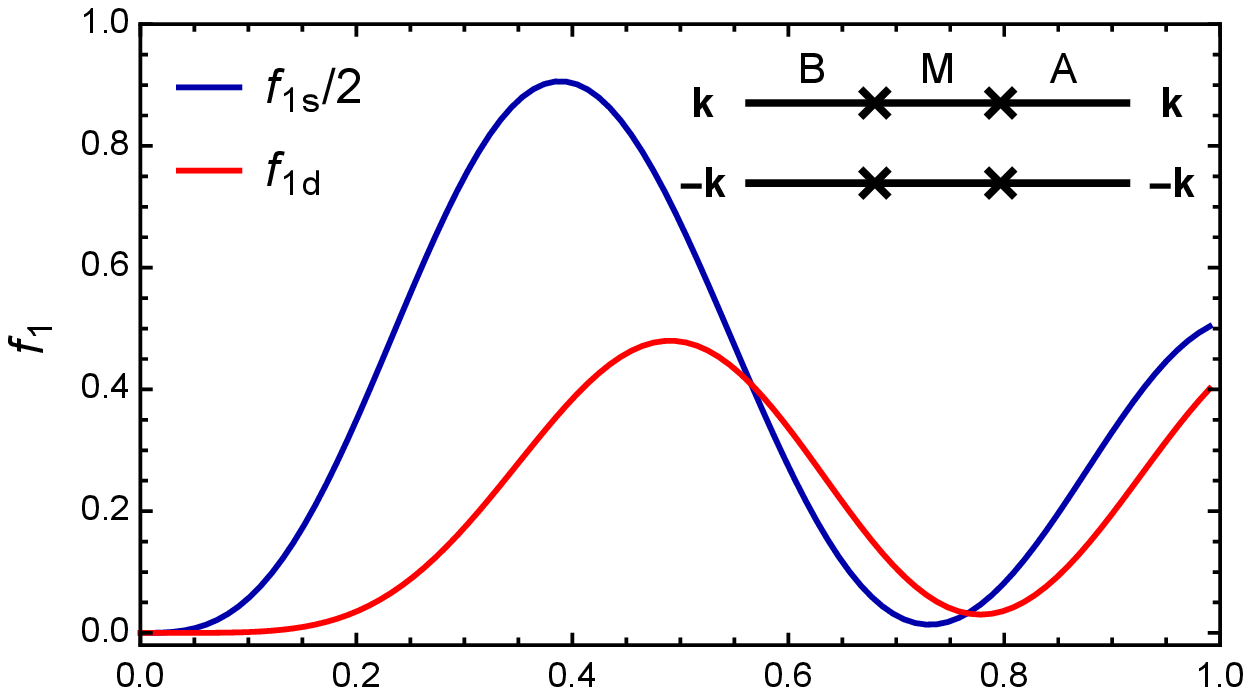}
\includegraphics[width=0.4\textwidth]{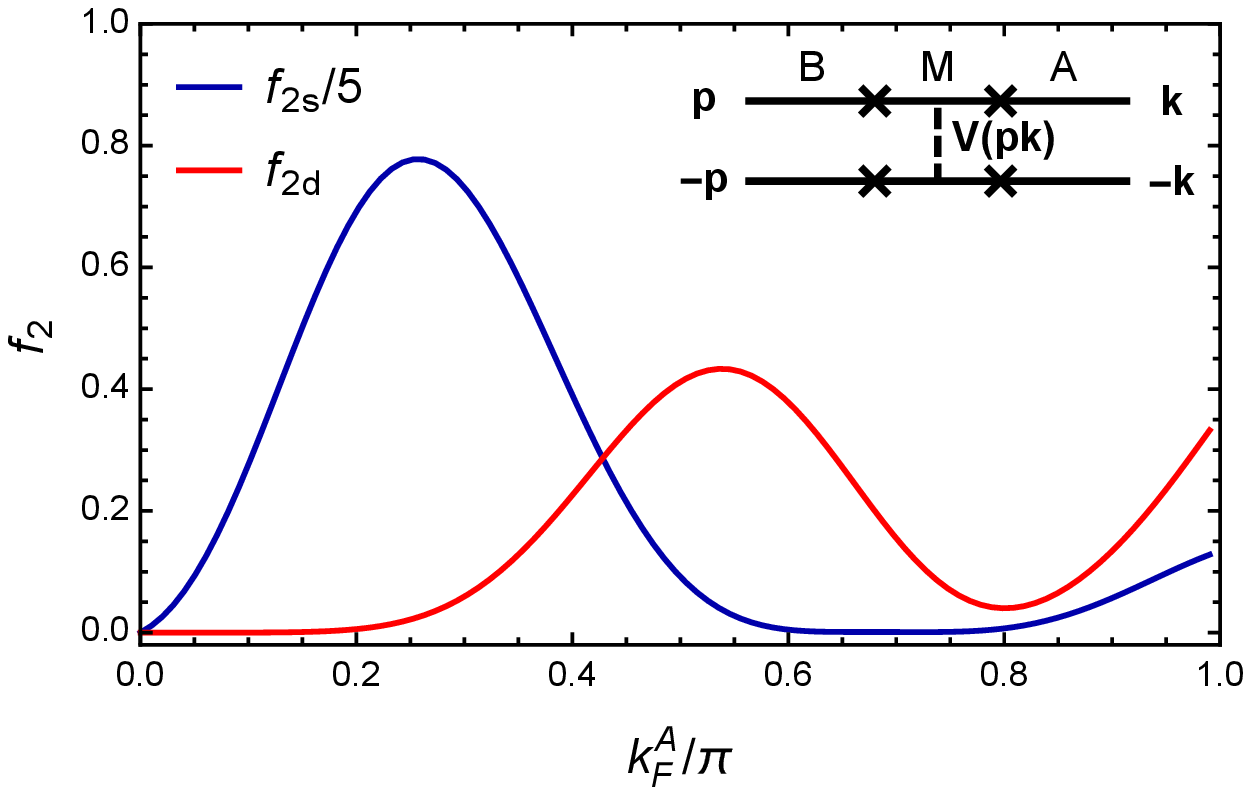}
\caption{\label{fig:fs} (Color online) Dependences of the $f_1$ (upper panel) and $f_2$ (lower panel) functions [Eqs. (\ref{eq:r1}) and (\ref{eq:r2})] on the Fermi circle radius in the $A$-layer for the cases of 
$s$- and $d$-wave SC in the $B$-layer. Note that the $s$-wave results $f_{1s}$ and $f_{2s}$ have been divided by 2 and 5, respectively. The corresponding single- and pair-tunneling processes are shown in the insets.}
\end{figure}

Inserting now the tunneling coefficients (\ref{eq:PQRS}) into Eq.~(\ref{eq:FPD}), 
we find a nonvanishing $F_{AB}$ in the $(\Delta^s_{B\kvec},\Gamma^-_2)$ and $(\Delta^d_{B\kvec},\Gamma^-_4)$ channels (as it was observed above), and evaluate the corresponding coupling constants $r$ using circular Fermi-surfaces for simplicity. The main contribution to $F_{AB}$ (\ref{eq:FPD}) stems from the region close to the Fermi surface in the $A$-layer. Away from nesting of the Fermi-circles in the $A$ and $B$ layers, the enhancement factor $r$ will be suppressed by  $\delta^{2}=[\xi_B(k^{A}_F)-\xi_A(k^{A}_F)]^{2}$, where $k^{A}_F$ is the Fermi circle radius in the $A$-layer. An estimate of the enhancement factor $r_1$ (due to single-particle tunneling considered so far) gives  
\begin{equation}
r_1\approx\frac{g_eg_o}{\delta^2}\frac{\Delta_0}{T_{C0}}\ln(\frac{\delta}{T_{C0}})f_1(k^{A}_F),
\label{eq:r1}
\end{equation}
where $f_1$ is a function that only depends on the Fermi-circle radius. Figure~\ref{fig:fs} shows $f_1(k^{A}_F)$ for the two combinations $(\Delta^s_{B\kvec},\Gamma^-_2)$ and $(\Delta^d_{B\kvec},\Gamma^-_4)$.  For representative values of $g_{e/o} \sim 0.1\delta$, $\Delta_0/T_{C0}\sim 30$, and $\Delta_0/\delta \sim 0.2$, we find $r_1\approx 1.5 f_1$.

Another process, that contributes to $F_{AB}$ in addition to (\ref{eq:FPD}), is the scattering of a Cooper pair of relative momentum $2{\bf p}$ in the $B$-layer to a pair of relative momentum $2\kvec$ in the $A$-layer, due to an electron-electron scattering potential of strength $V$ in the middle layer. 
This pair-tunneling process, sketched in the lower panel of Fig. \ref{fig:fs}, is not suppressed by the energy difference $\delta$ and gives the following contribution to the enhancement factor: 
\begin{equation}
r_2\approx \!-\frac{g_eg_o}{(\!\Delta\! E)^2}\frac{\Delta_0}{T_{C0}}N_0V\ln(\frac{2\omega_c}{\Delta_0})\ln(\frac{\omega_c}{T_{C0}})f_2(k^{A}_F,k^{B}_F), 
\label{eq:r2}
\end{equation}
where $N_0$ is the density of states and $\omega_c$ is an upper frequency cutoff. The function $f_2$ depends on the Fermi-circle radii in the $A$ and $B$ layers, and is plotted in Fig.~\ref{fig:fs} as a function of $k^{A}_F$ (at $k^{B}_F=k^{A}_F+\frac{\pi}{10}$).
Assuming $g_{e/o}\sim 0.1\Delta\! E$, $N_0|V|\sim 0.5$, and $\omega_c/\Delta_0\sim10$, we find $r_2\approx \pm 2.5f_2$. Depending on microscopics, $r_2$ is positive for an attractive potential 
(e.g., for a phonon and/or magnetically mediated interaction $V<0$), and negative for a repulsive one. From the above estimates, it seems plausible that
the single-particle and pair-tunneling processes can give a sizable enhancement factor of the order of
$|r| \sim 1$, as used in Fig. \ref{fig:ProxEnhancement}.
In addition, anti-ferromagnetic (AF) correlations are expected to arise 
in the iridate layer \cite{BJKim,Kim2012}.
Since pseudospins are spin-orbit composite objects,
their AF correlation is in fact a coherent mixture of real-spin
singlets and triplets, implying that the iridate AF correlations will
further facilitate a coherent singlet-triplet conversion.


In conclusion, we have shown how a coherent coupling
between a triplet and a singlet SC can be achieved by means of a time reversal
invariant conversion layer that effectively rotates singlet Cooper pairs into
triplets. The conversion is due to tunneling via the strong intraionic 
spin-orbit coupled states in the middle layer; a possible candidate for such a
''pair-convertor'' might be the iridium oxide Sr$_2$IrO$_4$. 
The coherent coupling leads to a dramatic enhancement of the triplet
superconductivity, existing well above its ''native'' critical
temperature $T_{C0}$. Experimentally, the enhanced triplet gap in the
quasiparticle spectrum and its temperature dependence as shown in Fig.\ref{fig:ProxEnhancement}
can be verified using ARPES and STM techniques. The proximity mechanism
considered here may also enable the stabilization 
of topologically non-trivial $p$-wave SCs.

We thank D. Scherer and P. Ostrovsky for helpful discussions. The work is supported by the Research Council of Norway, the Academy of Finland Center of Excellence program, the European Research Council (Grant No. 240362-Heattronics), and the Dutch Science Foundation NWO/FOM.



\begin{thebibliography}{99}
\bibitem{Mackenzie2003} A. P. Mackenzie and Y. Maeno, Rev. Mod. Phys. {\bf 75}, 657 (2003).
\bibitem{SigristUeda1991} M.~Sigrist and K.~Ueda, 
Rev. Mod. Phys. {\bf 63}, 239 (1991).
\bibitem{BergeretEtal2001} F.S.~Bergeret, A.F.~Volkov, and K.B.~Efetov,
Phys. Rev. Lett. {\bf 86}, 4096 (2001). 
\bibitem{Edelstein2003} V.M.~Edelstein, 
Phys. Rev. B. {\bf 67}, 020505(R) (2003).
\bibitem{EschrigLoefwander2008} M.~Eschrig and T.~L\"{o}fwander, 
Nat. Phys. {\bf 4}, 138 (2008).
\bibitem{TanakaSatoNagaosa2012} Y.~Tanaka, M.~Sato, and N.~Nagaosa, J. Phys. Soc. Jpn. {\bf 81}, 011013 (2012).
\bibitem{BergeretTokatly2013} F.S.~Bergeret and I.V.~Tokatly, 
Phys. Rev. Lett. {\bf 110}, 117003 (2013).
\bibitem{ParhizgarBlack-Schaffer2014} F.~Parhizgar and A.M.~Black-Schaffer, Phys. Rev. B {\bf 90}, 184517 (2014).
\bibitem{VortexPhase} A triplet order parameter can also be induced in a
  different type of system, such as 
in the vortex phase of high $T_C$ SCs  \cite{Lebed2006}.
\bibitem{Lebed2006} A.G.~Lebed, Phys. Rev. Lett. {\bf 96}, 037002 (2006). 
\bibitem{Sato09SO} M. Sato, Y. Takahashi, and S. Fujimoto,  
Phys. Rev. Lett. {\bf 103}, 020401 (2009).
\bibitem{Sau10} J.D.~Sau, R.M.~Lutchyn, S.~Tewari, and S.~Das Sarma, 
Phys. Rev. Lett. {\bf 104}, 040502 (2010).
\bibitem{NakosaiTanakaNagaosa2012} S.~Nakosai, Y.~Tanaka, and N.~Nagaosa, Phys. Rev. Lett. {\bf 108}, 147003 (2012).
\bibitem{Takei2013} S.~Takei, B.M.~Fregoso, V.~Galitski, and S.~Das Sarma, 
Phys. Rev. B {\bf 87}, 014504 (2013).
\bibitem{YangEtal2010} Z.~Yang, J.~Wang, and K.S.~Chan, J. Phys.: Condens. Matter {\bf 22}, 045302 (2010).
\bibitem{BrydonEtal2013} P.M.R.~Brydon, W.~Chen, Y.~Asano, and D.~Manske, Phys. Rev. B {\bf 88}, 054509 (2013).
\bibitem{BJKim} B.J.~Kim, H.~Ohsumi, T.~Komesu, S.~Sakai, T.~Morita, 
H.~Takagi, and T.~Arima, 
Science {\bf 323}, 1329 (2009). 
\bibitem{RashbaSOC} Due to breaking of mirror symmetry in the $z$-direction, 
a Rashba-type spin-orbit coupling can be induced at the $AB$-interface. Except
for very special and rare systems \cite{DiezEtal}, this effect is small and
therefore ignored here since it will not change the results qualitatively.
\bibitem{DiezEtal} M. Diez, A. M. R. V. L. Monteiro, G. Mattoni, E. Cobanera, T. Hyart, E. Mulazimoglu, N. Bovenzi,
C. W. J. Beenakker, and A. D. Caviglia, Phys. Rev. Lett. {\bf 115}, 016803 (2015).
\bibitem{AltlandSimons} A.~Altland and B.~Simons, 
{\it Condensed Matter Field Theory} (Cambridge University Press, Cambridge, U.K. 2010).
%
\bibitem{FNHigherOrder} Higher order terms in $H_{AB}$ can give rise to a phase dependent coupling between the two SCs even in the absence of a spin active interface, but these terms will in general be smaller and have a different dependence on the phase difference \cite{LuYip2009}.
\bibitem{LuYip2009} C.-K.~Lu and S.~Yip, Phys. Rev. B {\bf 80}, 024504 (2009).
\bibitem{Sato2009} M. Sato, Phys. Rev. B {\bf 79}, 214526 (2009).
\bibitem{QiEtAl2010} X.-L. Qi, T.L. Hughes, and S.-C. Zhang, Phys. Rev. B {\bf 81}, 134508 (2010).
\bibitem{FuBerg2010} L. Fu and E. Berg, Phys. Rev. Lett. {\bf 105}, 097001 (2010).
\bibitem{Leggett1975} A. J. Leggett, Rev. Mod. Phys. {\bf 47}, 331 (1975).
\bibitem{XuEtal1998} L. Xu, Z. Shu, and S. Wang, Phys. Rev. B {\bf 57}, 11654 (1998).
\bibitem{KadinGoldman1982} A.M.~Kadin and A.M.~Goldman, Phys. Rev. B {\bf 25}, 6701 (1982).
\bibitem{Hwang} H.Y.~Hwang, Y.~Iwasa, M.~Kawasaki, B.~Keimer, 
N.~Nagaosa, and Y.~Tokura, Nature Mater. {\bf 11}, 103 (2012).
\bibitem{footnote1} The pairing in ruthenates is in reality of multi-orbital 
nature and not yet fully understood 
(see, e.g., the recent works~\cite{Veenstra,ImaiWakabayashiSigrist2013,PuetterKee2012,ScaffidiRomersSimon2014}); however, our assumption of the 
$xy$ orbital is not of a principal importance for illustration of the 
basic idea of the present Rapid Communication. 
%
\bibitem{Veenstra} C.N.~Veenstra, Z.-H.~Zhu, M.~Raichle, B.M.~Ludbrook, 
A.~Nicolaou, B.~Slomski, G.~Landolt, S.~Kittaka, Y.~Maeno, J.H.~Dil, 
I.S.~Elfimov, M.W.~Haverkort, and A.~Damascelli, 
Phys. Rev. Lett. {\bf 112}, 127002 (2014).
\bibitem{ImaiWakabayashiSigrist2013} Y. Imai, K. Wakabayashi, and M. Sigrist,
Phys. Rev. B {\bf 88}, 144503 (2013).
\bibitem{PuetterKee2012} C.M. Puetter, H.-Y. Kee, EPL {\bf 98}, 27010 (2012).
\bibitem{ScaffidiRomersSimon2014} T.~Scaffidi, J.C.~Romers, and S.H.~Simon, Phys. Rev. B {\bf 89}, 220510(R) (2014).
%
\bibitem{SlaterKoster1954} J.C.~Slater and G.F.~Koster, 
Phys. Rev. {\bf 94}, 1498 (1954).
%
\bibitem{footnote2} Due to spin orbit coupling, the Ir $d_{x^2-y^2}$ and $d_{xy}$
orbitals mix, and will give an effective hopping between Cu $d_{x^2-y^2}$ 
orbitals in the $B$-layer and the Ir $d_{xy}$ orbitals. There is also a term corresponding 
to hopping in the straight $\evec_z$-direction. This effective term will 
not change the physics and is therefore ignored to keep the model as simple 
as possible.
%
\bibitem{Khaliullin} G.~Jackeli and G.~Khaliullin, 
Phys. Rev. Lett. {\bf 102}, 017205 (2009).
\bibitem{Kim2012} J. Kim, D. Casa, M. H. Upton, T. Gog, Y.-J. Kim, J. F. Mitchell, M. van Veenendaal, M. Daghofer, J. van den Brink, G. Khaliullin, and B. J. Kim, Phys. Rev. Lett. {\bf 108}, 177003 (2012).

\end{thebibliography}
\end{document}